\documentclass[prl,groupaddress,twocolumn,nofootinbib]{revtex4}

\usepackage{slashed}
\usepackage{braket}
\usepackage{graphicx}
\usepackage{amsmath}
\usepackage{amssymb}
\usepackage{epsfig}
\usepackage{hhline}
\usepackage{soul}
\usepackage{tabularx,pifont,multirow}
\usepackage{enumitem}
\usepackage{colordvi}
\usepackage{xcolor}

\newcommand{\be}{\begin{equation}}
\newcommand{\ee}{\end{equation}}
\newcommand{\bea}{\begin{eqnarray}}
\newcommand{\eea}{\end{eqnarray}}

\newcommand{\gev}{~{\rm GeV}}
\newcommand{\tev}{~{\rm TeV}}

\begin{document}
\preprint[\leftline{KCL-PH-TH/2020-42, CERN-TH-2020-131}

%

\title{ \bf
\boldmath
\Large    An Electroweak Monopole, Dirac Quantization \\ and the Weak Mixing Angle  } 


\author{John Ellis$^{~a,b}$, P.~Q. Hung$^{~c}$ and Nick E. Mavromatos$^{~a}$} 

\vspace{0.2cm}
\affiliation{$^a$Theoretical Particle Physics and Cosmology Group, Physics Department, King's College London, Strand, London WC2R 2LS, UK.
\vspace{0.25cm}\\
$^b$ Theoretical Physics Department, CERN, CH-1211 Gen\`eve 23, Switzerland; \\
National Institute of Chemical Physics \& Biophysics, R\"avala 10, 10143 Tallinn, Estonia
\vspace{0.25cm}\\
$^b$ Department of Physics, University of Virginia, Charlottesville, VA 22904-4714, USA}


\bigskip 

\begin{abstract}

\begin{center}
\textbf{Abstract }
\end{center}

We consider an extension of the Standard Model that was proposed recently by one of the current authors (PQH), which admits magnetic monopoles with a mass of order of a few TeV. We impose,  in addition to topological quantization in the SU(2) sector of the model, the Dirac Quantization Condition (DQC) required for consistency of the quantum  theory of a charged electron in the presence of the monopole. This leads to the prediction ${\rm sin}^2\theta_W = 1/4$, where $\theta_W$ is the weak mixing angle at the energy scale set by the monopole mass. A leading-order renormalization-group analysis yields the value of ${\rm sin}^2 \theta_W \simeq 0.231$ at the $Z$-boson mass, as measured by experiment, under suitable conditions on the spectrum of the extra particles in the model.

\end{abstract}
\maketitle

\section{Introduction}
\label{sec:intro}

The electroweak mixing angle $\theta_W$ is a free parameter within the Standard Model (SM) of particle physics. However,  it becomes possible to predict its value within extensions of the SM, e.g., by embedding the SM in a Grand Unified Theory (GUT), where the magnitude of $\theta_W$  is controlled by the details of unification~\cite{U1,U2,U3}, or in string theory~\cite{Raby:2007yc}. In SU(5) GUT theories, for instance, there is a characteristic tree-level prediction that
\be\label{su5GUT}
{\rm sin}^2\theta_W = {3}/{8}
\ee
at the GUT scale. This is renormalized by quantum loop effects in the SM that yield the prediction ${\rm sin}^2\theta_W \simeq 0.20$ at the $Z$-boson mass~\cite{U2}, which is close to, but different from, the experimental value ${\rm sin}^2\theta_W \simeq 0.231$ in the $\overline{MS}$ prescription~\cite{PDG}. The experimental value can be recovered by including the quantum corrections due to new particle degrees of freedom in the renomalization calculation. 
For example, including the supersymmetric partners of SM particles in the SU(5) GUT calculation reproduces very accurately the experimental value~\cite{susysin2}.

In this article we make a different prediction for ${\rm sin}^2\theta_W$ in an extension of the SM that is not a high-scale GUT, but rather a theory, proposed by one of the current authors (PQH) in \cite{hung}, that includes a topologically non-trivial magnetic monopole with a mass of a few TeV. This magnetic monopole is associated with a real scalar triplet of the SU(2) group, in a spirit similar to the Georgi-Glashow model~\cite{GG}, and obeys a topological quantization condition that stems from the known non-trivial homotopy properties of the SU(2) group. 

We show that this condition is not sufficient by itself to guarantee satisfaction of the Dirac Quantization Condition (DQC), \be\label{dqc1}
g_M \, e = \frac{m}{2}, \quad m \in \mathbb Z~,
\ee with $g_M$ the magnetic charge, which is required for consistency of the quantum theory of a charged particle such as the electron in the monopole's magnetic field~\cite{shnir,mm}. In the model of~\cite{hung} the DQC must be imposed as an extra condition~\cite{mskr}, which leads to the prediction
\be \label{PQH}
{\rm sin}^2 \theta_W = 1/4
\ee
at the monopole mass scale in the model.  This value is renormalized by extra particles with masses between $m_Z$ and the monopole mass that appear in the model, and the experimental value ${\rm sin}^2\theta_W \simeq 0.231$~\cite{PDG} is recovered under suitable conditions on the spectrum of these particles.

\section{Light Monopoles in the Model of~\cite{hung}
\label{sec:lightm}}

We now review briefly the main features of the model proposed in~\cite{hung}. It  involves {\em non-sterile} right-handed neutrinos with masses of the order of the electroweak scale, which participate in a seesaw mechanism for light neutrinos that is testable in principle at colliders,  e.g., by searching for like-sign dileptons with displaced vertices. For brevity, in what follows we term this model the EW-$\nu_R$ model. 

The central reason why the EW-$\nu_R$ model admits monopoles with masses at the electroweak scale, $\Lambda_{EW}$,  is 
that its right-handed neutrinos acquire~\cite{pqnur} electroweak-scale Majorana masses $M_R \propto \Lambda_{EW} \sim 246 \gev$ 
through their coupling to a {\em complex} Higgs triplet $\tilde{\chi}$. 

Because the $\nu_R$s are not sterile, consistency with the measured width of the $Z$-boson
requires $M_R \geq 46 \gev$, which implies $\langle \tilde{\chi} \rangle=v_M \propto \Lambda_{EW}$. Such non-sterile neutrinos would seriously affect the experimentally-verified relationship between the $W$- and $Z$-boson masses $M_W=M_Z \cos \theta_W$ in the SM, in the absence of an additional {real}  triplet of (Higgs-like) scalar fields $\xi$ with $\langle \xi \rangle = \langle \tilde{\chi} \rangle=v_M$~\cite{triplet}, 
which realize a 
custodial symmetry in the EW-$\nu_R$ model~\cite{pqnur}. The $\xi$ triplet is hypercharge-neutral and gives rise to a {\it finite-energy} electroweak monopole solution of the classical Euler-Lagrange equations of the model~\cite{hung}, following the pattern of the SO(3) monopole in the Georgi-Glashow model~\cite{GG}, discovered by `t Hooft~\cite{thooft} and Polyakov~\cite{polyakov} (the 'tHP monopole). In that model the electromagnetic U$_{\rm em}$(1) gauge group is embedded into the SO(3) gauge group, whose algebra is isomorphic to that of the SU(2) appearing in the model of \cite{pqnur,hung}. However, in the model of~\cite{hung}  U$_{\rm em}$(1) is a combination of the SU(2) and U$_{\rm Y}$(1) of the SM, parametrized by $\sin^2 \theta_W$ in the usual way.

We now review the topological arguments~\cite{hung} for the existence of the monopole, clarifying the independence of the 
topological quantization condition from the DQC that we explore subsequently. The EW-$\nu_R$ model 
contains~\cite{hung,pqnur}. In addition to the real triplet $\xi$ and the complex triplet $\tilde{\chi}$, four complex Higgs doublets, $\phi^{SM}_{i}$  (which couple to SM fermions only), and $\phi^{M}_{i} $(which couple to  mirror fermions ($MF$)~\cite{Mirror}, each with $i=1,2$ 
and some Higgs singlets $\phi_S$ that are not relevant to the magnetic monopole solution. 
The vacuum alignment that guarantees the custodial electroweak symmetry has $\langle \tilde{\chi} \rangle= \langle \xi  \rangle= v_M$ \cite{pqnur}. The vacuum manifold of the Higgs sector is 
\be
\label{vacuum}
S_{vac} = S^{2} \times S^{5} \times \sum_{i=1,2} S^{3}_{SM_i} \times \sum_{i=1,2} S^{3}_{M_i}  \,.  
\ee  
where an $n$-sphere $S^n$ is described by the equation $x_1^2 +x_2^2 +..+x_{n+1}^2 =$ constant. 
Here, the $x_i$ denote the various scalar field values, and the constant radii of the various spheres correspond to the vacuum expectation values of the various Higgs field components. 
The second homotopy group of the vacuum manifold of the EW-$\nu_R$ model (\ref{vacuum}) is therefore~\cite{hung}
\bea
\label{result}
\pi_2(S_{vac})&=&\pi_2(S^2)  \oplus \pi_2(S^5)  \oplus_{i=1,2} \pi_2(S^{3}_{SM_i,M_i}) \\ \nonumber
                       &=&  \pi_2(S^2) = Z \, ,
\eea
which is the standard topological argument for the existence of an 'tHP monopole~\cite{thooft,polyakov}. We see that the EW-$\nu_R$ model
has a topologically-stable monopole solution thanks to the real SU(2) triplet $\xi$, corresponding to the sphere $S^2$, for which 
$\pi_2(S^2) = Z$. 
Thus the EW-$\nu_R$ model makes an interesting connection between the light neutrino masses and the existence of magnetic monopole solutions. 
  
It was noted in \cite{hung} that, since $S^2$ is associated with the vacuum manifold of the real triplet $\xi$, topological quantization would involve the SU(2) coupling $g$, rather than the electromagnetic coupling $e$, leading to the following quantization condition for the magnetic charge $\tilde{g}$ of the monopole:
\be
\label{quantize}
\frac{g \tilde{g}}{\hbar \, c}=n \,, \quad n \in \mathbb Z \, . 
\ee          
From now on we work in units with $\hbar =c=1$.            
The fact that the quantization condition \eqref{quantize} is in terms of the monopole charge $\tilde{g}$ {and}  the {weak charge} $g$ instead of the  electric charge $e$ appearing in the standard DQC is a characteristic feature of the model of \cite{hung}. It distinguishes the monopole in the model of~\cite{hung} from the 'tHP magnetic monopole or the Cho-Maison monopole~\cite{cho-maison} and its finite energy extensions~\cite{cho-maison2}, to which the standard DQC applies.

It is easy to see that the condition \eqref{quantize} 
follows from the 
equation of motion for the field strength of the $W_3$ SU(2) gauge boson, 
 $W_{3}^{\mu \nu}=\partial^{\mu} W_{3}^{\nu}-\partial^{\nu} W_{3}^{\mu} + \frac{1}{v_{M}^{3} g} \varepsilon_{abc} \xi^{a} \partial^{\mu} \xi^{b} \partial^{\nu} \xi^{c} $:
\be\label{eqmot}
\partial_{\mu } \tilde{W_{3}}^{\mu \nu} \; = \; K^{\nu} \, ,
\ee
where $\tilde{W_{3}}^{\mu \nu} \equiv \frac{1}{2} \varepsilon_{\mu \nu \sigma \rho}W_{3}^{\sigma \rho}$ and $K_{\mu} \equiv{} \frac{1}{2} \epsilon_{\mu \nu \sigma \rho} \partial^{\nu} W_{3}^{\sigma \rho}$ is a topological current that is automatically conserved by definition. The {\em topological charge}, $g_M$, is defined as $g_{M}=\int d^{3}x K_{0}$. Carrying out the integration, one obtains the topological quantization condition \eqref{quantize}.

Including the full electroweak gauge group structure 
SU(2)$\times$U$_{\rm Y}$(1), which is broken 
down to the electromagnetic U$_{\rm em}$(1) by the complex Higgs doublets 
and the triplet $\tilde \chi$ of the EW-$\nu_R$ model~\cite{hung}, one sees that
the $W_{\mu}^{3}$ gauge field of the SU(2) subgroup is a mixture of  the $Z$-boson and photon 
fields, parametrized as usual by the weak mixing angle $\theta_W$:
$W_{\mu}^{3}= \cos \theta_W Z_{\mu} + \sin \theta_W A_{\mu}$, with $\sin \theta_W = g'/\sqrt{g^2 + g'^2}$ where $g'$ is the U$_{\rm Y}$(1)
coupling.
It should be stressed that, under the breaking $U(1)_W \times U(1)_Y$, corresponding to some v.e.v. $v_2$, 
which symbolically represents the contribution of the other Higgs fields of the model, 
the $W_\mu^3$ is no longer a mass eigenstate, but the $Z_\mu$ and $A_\mu$ are the correct, physical, {\em mass eigenstates}.

The corresponding field strengths are
\be
\label{W3new}
W^{3}_{ij} =  \cos \theta_W Z_{ij} + \sin \theta_W F_{ij} \,,
\ee  
where $F_{ij}$ is the usual electromagnetic field-strength tensor and $Z_{ij}$ is the $Z$ field-strength tensor. 
This mixing between the photon and the $Z$-boson is the reason why the terminology ``$\gamma$-Z magnetic monopole" 
was used in \cite{hung} to describe the magnetic monopole solution.

As discussed in \cite{hung}, the magnetic monopole has a mass
\be\label{monmass}
M_M=\frac{4\pi v_M}{g} f(\lambda/g^2) \,,
\ee
where the function $f(\lambda/g^2)$ varies between 1 for $\lambda=0$ (the Prasad-Sommerfield limit~\cite{PS})
and 1.78 for $\lambda=\infty$. 

The phenomenological analysis of Ref.~\cite{pqnur} shows that the value of $v_M$ is 
bounded from below by the $Z$ width (assuming only three light neutrinos): $v_M > M_Z/2 \sim 45.5 \gev$, 
and from above by the sum of the squared scalar fields VEVs in the model: $(\sum_{i=1,2} v_i^2 + v^{M,2}_{i}) + 8 v_M^2 = (246 \gev)^2$. 
The monopole mass range given by \eqref{monmass}  is then obtained by saturating the bounds on $v_M$: 
\be\label{monmass2}
M_M \; \sim \; 890 \gev - 3.0 \tev\,,
\ee
The corresponding magnetic field intensity arising in the breaking of $\rm SU(2) \times U_Y(1) \to U_{\rm em}(1)$ is defined by~\cite{hung} 
\be\label{magint}
B^{\gamma Z}_{i} = - \frac{1}{2} \epsilon^{ijk} \, W_{jk}^3 = \frac{\widehat r_i}{g\, r^2}, 
\ee
where $\epsilon^{ijk}$ ($i,j,k=1,2,3$) is 
the totally antisymmetric symbol in three Euclidean (spatial) dimensions. 
We then obtain from \eqref{W3new}~\cite{hung}
\bea
B^{\gamma Z}_{i}&=&  \cos \theta_W B^{Z}_{i} + \sin \theta_W B^{\gamma}_{i} \nonumber \\
&=& \frac{1}{g r^2} \hat{r}_{i} ( \cos \theta_W e^{-M_{Z} r} + \sin \theta_W ) \nonumber \\ 
                            &=& \frac{\sin \theta_W}{e r^2} \hat{r}_{i} ( \cos \theta_W e^{-M_{Z} r} + \sin \theta_W ) \,,
\label{gammaZ}
\eea  
where $B^{Z}_{i} =\frac{1}{g r^2} \hat{r}_{i} \exp^{-M_Z r}$ and $B_i^{\gamma}=\frac{1}{g r^2} \hat{r}_{i}$ are the short-range $Z$-magnetic field and the long-range magnetic field, respectively, and 
\be\label{egrel} 
e = g \sin \theta_W
\ee
denotes the usual electromagnetic coupling, as in the SM. 
We observe that in the limit $v_2=0$ and $g'=0$, i.e., $\theta_W=0$ and $M_Z=0$, one recovers \eqref{magint} from \eqref{gammaZ}, confirming the mathematical correctness and consistency of the latter expression.

We note the exponential damping factor
$\propto \exp(-M_{Z} r)$  in the expression (\ref{gammaZ}) for the magnetic field strength, due to the 
finite $Z$-boson mass, $M_Z \ne 0$. The short- and long-range parts of $B^{\gamma Z}_{i}$ become comparable in strength at a distance 
$r= \frac{1}{M_Z} \ln (\cot \theta_W) \sim 0.6/M_Z$ from the centre of the monopole, which is well inside its core. At large distances compared to the monopole core radius, $r \gg R_c \sim (gv_M)^{-1} $, the magnetic field differs in strength from that of a point-like Dirac monopole by a factor $\sin^{2} \theta_W$.

At these large distances, the $\gamma-Z$ magnetic field  is 
\be
\label{Bnew}
B^{\gamma Z}_{i}  \approx  \frac{\sin^{2} \theta_W}{e r^2} \hat{r}_{i} \,.
\ee    
The true magnetic field, $B_i$, $i=1,2,3$, is defined in terms of the electromagnetic tensor $F_{ij}$, which is seen from \eqref{W3new} to be related
to $B^{\gamma Z}_{i}$ by a factor of $1/\sin \theta_W$, so that at large distances compared to the core monopole $R_c\sim (gv_M)^{-1}$:
\be\label{truemagf}
B_{i}  \approx  \frac{\sin \theta_W}{e r^2} \hat{r}_{i} \, \quad  i=1,2,3\,, \quad r \gg R_c
\ee   
Comparing this magnetic field~\cite{hung} with the conventional definition of the magnetic charge of the monopole~\cite{shnir}, we see that
\be\label{elmagch}
g_M = \frac{{\sin} \theta_W}{e} \, .
\ee
Eq.~\eqref{elmagch} can be understood from the definition of the magnetic charge, $g_M$, which is topological in our model, 
and given by the topological quantization condition, Eq.~\eqref{quantize}. Taking  $n=1$ and setting $\tilde g = g_M$ in that relation,  one obtains: $\tilde g = g_M =1/g$. 
Using the standard form of the magnetic field of a magnetic monopole, far away from its centre (placed at the origin)~\cite{shnir} 
\be\label{magfmon}
B_i^\gamma = \frac{g_M}{r^2} \, \hat r_i =\frac{1}{g\, r^2}\, \hat r_i \, ,
\end{equation}
and using \eqref{egrel}) we find
\begin{equation}
    B_i^\gamma = \frac{{\rm sin}\theta_W}{e\,r^2} \, \hat r_i~.
\ee
We thus recover the expression \eqref{truemagf} for the effective magnetic field  in our model, defined via \eqref{W3new}, at distances larger than the monopole core radius. We stress again that the topological magnetic charge is {\it not} of the form ${\rm sin}^2\theta_W/e$, which appears as the coefficient of $1/r^2$ in the expression for the $\gamma-Z$ ``SU(2) magnetic'' field $B_i^{\gamma Z}$, but of the form \eqref{elmagch}, carrying a single power of ${\rm sin} \, \theta_W$, when expressed in terms of the electron charge $e$ in the model.

We stress that the above solution for the $\gamma-Z$ magnetic field respects rotational symmetry, 
and takes the Dirac form at large distances, but vanishes at the centre of the monopole, so that the solution
has finite energy~\cite{hung}. 
In this respect, our solution differs from the (dumbbell) monopole of Nambu in the conventional SU(2)$\times$U(1)$_Y$
Standard Model~\cite{nambu}, which breaks spontaneously the rotational (and hence Lorentz) symmetry, due to the 
presence of a $Z$-flux string. The latter is also responsible for the confinement of the Nambu monopole with its 
antimonopole. As demonstrated in \cite{vachaspati}, the presence of the $Z$-string in that case leads to the 
satisfaction of the DQC without any restriction on the weak mixing angle, unlike our situation as discussed below. 
Specifically, in the Nambu electroweak monopole, the (non-singular) hypercharge U(1)$_Y$ magnetic field 
$\vec B_{{\rm U(1)}_Y}$ emanating from the monopole is compensated by that entering the monopole via the $Z$-string,
leading to a divergence-free component of the field, $\vec{\nabla}\cdot \vec B_{{\rm U(1)}_Y} =0$.
As a consequence the Nambu monopole pertains exclusively to the SU(2) sector of the Standard Model, 
which contains only the coupling $g$ of that sector, and obeys the standard topological quantization
condition, without any factors of sin$\theta_W$. This is not the case in the spherically-symmetric magnetic $\gamma-Z$ field \eqref{magfmon} in the model of \cite{hung} which, we stress again, 
does not possess any such $Z$-flux string. 
We stress, though, that, as in the non-singular dumbbell solution of \cite{nambu}, 
the total energy of the monopole solution of \cite{hung} is finite, leading to a finite mass for the monopole of 
order a TeV, as discussed in \cite{hung} and here.
 
\section{The Dirac Quantization Condition}
\label{sec:DQC}

As a consequence of \eqref{elmagch},  {\it the DQC \eqref{dqc1} is in general violated by the weak mixing-angle factor in $g_M$}. 
Thus the electron wave function would {\it not} be single-valued along a loop that surrounds the monopole
at large distances from the monopole centre~\cite{shnir}. 

The DQC is often derived by considering the translation of an electron
around a Dirac string of the type connected to a point-like monopole.
However, in the current monopole solution, there is no Dirac string, hence the reader might think that the topological quantization \eqref{quantize}, which stems from the non-trivial homotopy structure \eqref{result}, is sufficient
for the consistency of the model. However, the DQC is a general condition, derivable from
consistency conditions far away from the monopole centre, independent of the details of the monopole solution and whether it has an attached string. Hence, for the consistency of a theory, the DQC must always be imposed, if not automatically satisfied as is the case of the 'tHP monopole. Below we outline several arguments showing how the DQC emerges, independent of the details of the monopole structure.

First, it can be shown~\cite{shnir,mm}  that the DQC corresponds to the the {\it quantization condition of the angular momentum} of a {\it classical} charged particle, say an electron for concreteness, moving in the background of a magnetic pole at rest. Thomson~\cite{thomson} considered such a system twenty-seven years before Dirac's theory of magnetic monopoles, and showed that the total classical angular momentum of the electron  in this system is
\bea\label{angL}
\vec L =  m  \vec r  \times  \frac{d \vec r}{d t} - e g  \frac{\vec r}{r}~,
\eea
where the second term on the right-hand side is the contribution due to the interaction of the electron with the magnetic monopole field, as derived from the pertinent Poynting vector. The DQC \eqref{dqc1} follows from \eqref{angL} as a consequence of the usual quantization rule of the angular momentum, which is required to take on integer or half-integer values, in the case where both the electron and the monopole are at rest ($d\vec r/dt=0$).

Secondly, in the context of the monopole of the model of \cite{hung}, the DQC can be understood~\cite{mm} by making an analogy between this monopole solution, which crucially involves a Higgs breaking SU(2)$\times$U$_{\rm Y}$(1) $\rightarrow$ U$_{\rm EM}$(1) (in contrast to the conventional 'tHP monopole~\cite{thooft,polyakov} that is based on a Higgs breaking the simple group  SU(2) $\rightarrow$ U(1) as in the Georgi-Glashow model~\cite{GG}) with an Abelian Wu-Yang (WY) monopole~\cite{wu}, which does not involve a Dirac string. 

In such a case, it is well known~\cite{mm} that the DQC can be derived by covering the 3-space surrounding the WY monopole at the origin by two hemispheres (North (N) and South (S)) and considering  
a closed loop $\ell$ that lies entirely in the ``equator region'' in which the two hemispheres overlap. The loop can be located far away from the centre of the monopole. The wave function of an electrically-charged particle circulating the loop, say an electron of charge $e$, will pick up a  
phase $e \oint_\ell d\ell \cdot \mathbf A^{\rm S,N} $, where $\mathbf A^{N\, (S)}_\mu$ denotes the electromagnetic potential in the North (South) hemisphere. Applying Stokes's  theorem in each hemisphere, we may write:
\bea
e \oint_\ell d\ell \cdot \mathbf A^{\rm N} &=& e \int_{{\mathbb R}^{\rm N}}\, d{\mathbf S} \cdot (\nabla \times \mathbf A^{\rm N}) = 
e \int_{{\mathbb R}^{\rm N}} \, d{\mathbf S} \cdot \mathbf B, \nonumber \\
e \oint_\ell d\ell \cdot \mathbf A^{\rm S} &=&-e \int_{{\mathbb R}^{\rm S}} \, d{\mathbf S} \cdot (\nabla \times \mathbf A^{\rm S}) = 
-e \int_{{\mathbb R}^{\rm S}} \, d{\mathbf S} \cdot \mathbf B, \nonumber \\
\eea
where $\mathbf B$ is the magnetic field of the monopole, whose asymptotic structure is given by $ \mathbf B \sim \frac{g_M}{r^2} \,\hat{\mathbf{r}}$, with $g_M$ the magnetic charge. The action $S$ is therefore defined up to a term
\be\label{actionWY}
\Delta S = e \int_{\mathbb R^{\rm N} \cup \, \mathbb R^{\rm S}} d{\mathbf S} \cdot \mathbf B = \int_V\, d^3 r \, \nabla \cdot \mathbf B = 4\pi e  g_M,
\ee
where we used Gauss's law over the entire space volume $V$, and the corresponding Maxwell's equations for the monopole magnetic field.  The requirement that the action change \eqref{actionWY} does not affect any physical observables implies the DQC, in a way independent of the topological argument \eqref{result}.~\footnote{A similar result is obtained if one considers, alternatively, the gauge dependence of the electron wave function in a monopole field~\cite{shnir}.} 

We stress that the use of a non-compact Abelian U(1) gauge field in the above argument is specific to the fact that the model contains the Standard Model group with its usual breaking to U(1)$_{\rm}$, which plays the role of the non-compact Abelian group associated with the gauge potential used above. One cannot apply this argument to the 
standard 'tHP monopole, for which the topological quantization of the non-Abelian simple group SU(2) replaces the DQC.

This is the central point of this article: in the model of \cite{hung}, unlike the 'tHP monopole, 
{\it the topological quantization rule} (\ref{quantize}) stemming from the homotopy properties of the SU(2) group
is {\it not sufficient for the quantum consistency} of the electron wave function in the presence of the magnetic field 
induced by the $\gamma-Z $ monopole. 

In a similar spirit to the Kalb-Ramond monopole of \cite{mskr}, 
one must impose the DQC as an {\it additional} constraint:
\be\label{DQChung}
e \, g_M  = \frac{m}{2}, \quad m \in \mathbb Z \, .
\ee
We then obtain from \eqref{DQChung} and \eqref{elmagch} a consistency condiion for the 
weak mixing angle, and the prediction
\be\label{pred2}
{\sin} \theta_W = \frac{m}{2}\, \, \Rightarrow \, \, 
{\sin}^2 \theta_W = \frac{m^2}{4}\,,  \, m \in \mathbb Z \, ,
\ee
where  $\sin^2 \theta_W$ is the quantity that it is usually quoted in experimental measurements~\cite{Alternative}.
Since ${\rm sin}\theta_W \le 1$, the condition (\ref{pred2}) allows only two topological sectors, namely $|m|=1$ and $|m|=2$. 
The case $m=2$ would imply $\sin \theta_W=1$, which corresponds to the limit $g/g' \to 0$ and a massless $W$ boson. 
In the allowed case $m=1$ we have the prediction
\be\label{pred}
{\rm sin}^2\theta_W = \frac{1}{4} \,,
\ee
which is close to the experimental value $\sin^2 \theta_W \simeq 0.231$.

At this point we would like to offer further independent support to the restriction \eqref{pred2} by considering the coherent-state approach to the composite monopoles proposed in \cite{drukier}, which also lead to the arguments on the impossibility of producing composite monopole-antimonopole pairs at colliders, as a result of  the extreme suppression of the pertinent production cross sections of order $e^{-4\pi/\alpha}$, with $\alpha$ the electromagnetic fine structure constant.  

\section{Coherent-State Approach}
\label{sec:coherent}

According to the qualitative arguments of \cite{drukier}, a composite monopole state, such as the 'tHP monopole~\cite{thooft,polyakov} or that in the model of \cite{hung} that we study here, consists of $\overline n_{\rm quanta} V_c $ coherent states of Higgs and gauge quanta, where $V_c$ is the volume of a sphere of radius equal to the radius of the monopole core $R_c$, and $\overline n_{\rm quanta}$ denotes the density of the Higgs or gauge quanta (these number densities are of the same order of magnitude, as argued in \cite{drukier}). The quanta corresponding to the electrically-charged gauge states ($W$) with charge $e$, couple to the electromagnetic photon with a collective coupling 
\begin{equation}\label{collcoupl}
g^\gamma_{\rm coll} \sim \overline n_{\rm quanta} \, e, 
\end{equation}
Following the arguments of \cite{drukier} for our case,  the number density of the extra scalar triplet Higgs quanta is
\begin{equation}
\overline n_{\rm quanta} \sim m_{``H"} v_M^2 \,.
\end{equation}
where $m_{``H"}$  is the mass of the appropriate Higgs field in the model of \cite{hung}, and $v_M$ its 
vacuum expectation value, as reviewed above. 
Thus, the total number of Higgs quanta is
\begin{equation}\label{higgs}
\overline n_{\rm quanta}= (\frac{4}{3} \pi R^3_c) \, m_{``H"} v_M^2 \,.
\end{equation}
where the monopole core radius is given by $R_c \sim (g v_M)^{-1}$, $m_{``H"} \sim \sqrt{\lambda} \, v_M$, and $\lambda$ is the corresponding Higgs-self-interaction coupling. 
Assuming following \cite{drukier} that $\sqrt{\lambda}/3 \sim g$ in order of magnitude, so that $m_{``H"} \sim m_W \sim g v_M$,  one then obtains
\begin{equation}\label{nmH}
\overline n_{\rm quanta} \sim (\frac{4}{3} \pi) \frac{\sqrt{\lambda}}{g^3} \sim \frac{1}{\alpha_2} \end{equation}
in order of magnitude, where, $\alpha_2 \equiv g^2/(4\pi)$ is the fine structure constant of the SU(2) group.

In the pure SU(2) 'tHP-monopole case studied in \cite{drukier}, the electric charge of the $W$ states is $g$, since there is no mixing, hence the collective coupling to photons ($\sim W_\mu^3$) would in this case be given by 
\eqref{collcoupl} but with $e$ replaced by $g$. In this case, \eqref{higgs} implies the topological charge quantization \eqref{quantize}, with $n=1$, upon identifying $g_{\rm coll}^\gamma \sim 4\pi \, \tilde g$, where $\tilde g=1/g$. 

However, in the model of \cite{hung}, due to the presence of the Standard-Model group  SU(2)$\times$U$_{\rm Y}$(1) breaking down to the non-compact electromagnetic group U$_{\rm EM}$(1), and thus a non-trivial weak mixing angle $\theta_W \ne 0 \, {\rm mod}\,\pi$, we have $g=e/\sin \theta_W$,  with $e$ the electric charge that would couple the charged $W$ states to photons. Hence in that case, we obtain from \eqref{nmH}
\be\label{nmH2}
\overline n_{\rm quanta} \sim \frac{\sin^2 \theta_W}{\alpha^2} \, ,   
\ee
where $\alpha=e^2/(4\pi)$ is the fine structure constant of electromagnetism. 
The reader should notice the dependence of $\overline n_{\rm quanta} $ on the weak mixing angle, in contrast to the case of \cite{drukier}, based on the conventional 'tHP monopole, for which 
$ \overline n_{\rm quanta} \sim \frac{1}{\alpha}$. This feature of the monopole is exclusive to the model of \cite{hung}. 
Upon recalling~\cite{drukier} that the number \eqref{higgs} is also of the same order of magnitude as the number of charged gauge quanta $W^{\pm}$,  each of which carries a charge $e$, then we obtain 
the total charge, i.e., the collective coupling \eqref{collcoupl} of the monopole to photons, in our model:  
 \begin{equation}\label{gcoll2}
g^\gamma_{\rm coll} = \overline n_{\rm quanta} \, e \sim   \frac{\sin^2 \theta_W}{\alpha} \, e \, .
 \end{equation}
As discussed previously ({\it c.f.} \eqref{Bnew}), the $B^{\gamma Z}$ magnetic field corresponds to a magnetic charge $Q^{\gamma Z}=\frac{\sin^2 \theta_W}{e} \sim  g_{\rm coll}^\gamma$, in order of magnitude. However, $B^{\gamma Z} = \sin \theta_W B^{\gamma} + \cos \theta_W B^{Z}$. So the true magnetic field $B^{\gamma}$ corresponds to a magnetic charge $g_M = Q^{\gamma Z}/\sin \theta_W = \frac{\sin \theta_W}{e}$ ({\it c.f.} \eqref{truemagf}). In this case, the collective coupling \eqref{gcoll2} is linked to $g_M$ via $g_{\rm coll}^\gamma \sim 4\pi\,g_M\, {\rm sin\theta_W}$.

Far from the core, this monopole behaves like a Dirac monopole and the DQC applies: $e g_M=m/2$, $m \in \mathbb Z$. Upon comparing with the above result \eqref{gcoll2}, we obtain \eqref{pred}, i.e., $\sin^2 \theta_W=1/4$.

From the above discussion it follows that the topological charge $\tilde g$ of the 'tHP or the monopole of \cite{hung}, and in general of non-Abelian composite monopoles characterised by groups with non-trivial homotopy leading to a topological quantization rule \eqref{quantize}, can be identified with the collective coupling of the constituent coherent-state charged gauge quanta to photons \eqref{collcoupl}~\cite{drukier}. 

We speculate that a rigorous proof of this relation might be provided by extending the (1+1)-dimensional prototype study of \cite{dvali} to four-dimensional composite monopoles, of finite energy, including that of \cite{hung} studied here,  which are topological solitons. It was recognised in \cite{dvali}, considering a quantum coherent state of a (1+1)-dimensional topological soliton $|\rm soliton\rangle$ as a tensor product state of an infinity of constituent coherent states $|\alpha_\kappa \rangle$ corresponding to momentum $k$, 
$$|\rm soliton\rangle = \prod_{\otimes \, k} \, |\alpha_\kappa \rangle, $$
that the topological charge arises from the Noether charges of the microscopic constituent coherent states, which thus explains the conservation of the topological charge from basic properties of the constituent coherent states. Specifically, the topological charge arises from an infinite occupation number of zero momentum quanta, which in the (1+1)-dimensional model flow in one direction.
In the composite monopole case, the Noether charge is the electric charge, whilst the topological (magnetic) charge is the collective coupling \eqref{collcoupl} of the coherent constituent $W$ quanta to photons. 

\section{Absence of Confinement}
\label{sec:nonconf}

Some important comments are in order at this point, concerning the relation \eqref{DQChung}, from which our main result \eqref{pred2} follows. The implicit assumption underlying such a relation is that free monopoles exist after spontaneous symmetry breaking in the model. In fact, one could have thought of avoiding the imposition of the DQC by 
considering monopoles either confined with their antimonopoles by magnetic flux strings, in which case there is no constraint between magnetic and electric charges, or confined in groups of a certain fixed number, which would lead to a much weaker quantisation condition than \eqref{DQChung}. Such cases are discussed, for instance, in the review of \cite{preskill}. 

In Section 5.1 of that work, the author discuses magnetic monopoles arising from a toy model characterised by the following pattern of spontaneous gauge-symmetry breaking:
SU(3)$\stackrel{v_1}{\longrightarrow}$SU(2)$\times$U(1)$\stackrel{v_2}{\longrightarrow}$U(1), via appropriate Higgs vacuum expectation values (VEVs) with $v_2 \ll v_1$.  
As discussed in \cite{preskill}, the monopoles and anti-monopoles of mass $v_1/e$ from SU(3)/SU(2)$\times$U(1) are bound to each other by flux tubes and do not survive below $v_1$. 
Only the lighter ones of mass $v_2/e$ survive in that model. 

This is a quite different scenario from the electroweak monopole model of \cite{hung} discussed here. The VEV of the real triplet $v_M$ is comparable in value to the other VEVs (doublets, complex triplet),
so the hierarchical scenario discussed in \cite{preskill} does not apply. Moreover, as indicated above, in the paragraph following \eqref{egrel}, the short-range $Z$-magnetic field becomes subdominant 
relative to the long-range magnetic field already inside the core of size $R \sim (g v_M)^{-1}$ and, outside the core, only the long-range true magnetic field is present. 
These considerations imply that the monopoles are not confined with their antimonopoles, and hence the condition \eqref{DQChung} applies. 
There are no long-range magnetic fluxes binding a monopole with an anti-monopole in our model, due to the short-range nature of the $Z$-magnetic field, unlike the example discussed above. 
There, the breaking pattern is $H_1 = $SU(2)$\times$U(1)$\to H_2 = $U(1) at $v_2$, where $H_2 \in $SU(2) and the U(1) magnetic fluxes are confined, 
binding the heavy monopoles and anti-monopoles of mass $v_1/e$.

\section{Renormalization Effects}
\label{sec:renzn}

The DQC \eqref{DQChung} is a discrete consistency condition that should be understood as applying to the electric charge
in the large-distance (IR) limit and the monopole charge measured at the monopole mass \eqref{monmass}.
There is no renormalization of the monopole charge below this scale, as there are no magnetically-charged objects with masses
below \eqref{monmass}. On the other hand, as $\sin^2 \theta_W$ is related to the SU(2) and U(1)$_{\rm Y}$ couplings, 
it is in general subject to scale-dependent renormalization in the non-magnetic sector where experiments are performed.
This is a well-understood effect that has been studied in detail in many GUT models such as SU(5)~\cite{U2, susysin2}.

We have made leading-logarithmic one-loop calculations of the renormalization of ${\sin}^2 \theta_W$ 
from the monopole mass scale $M_M$ down to the $Z$-boson mass $M_Z$ for different values of $M_M$, the numbers of light families $F$ (including both SM and mirror fermions), light Higgs doublets $n_H$, 
real triplets $n_3$ and complex triplets $\bar{n}_3$ with masses
below $M_M$ that enter the evolution. We use the notation $x_W \equiv {\sin}^2\theta_W(M_Z^2)$
and assume that, at $M_M$, ${\sin}^2\theta_W(M_M^2) = g'^2/[g^2+g'^2]=1/4$ giving $\alpha^\prime =(1/3) \alpha_2$, and the following
one-loop renormalization formula
\bea
\label{xW}
\hspace{-1mm}
x_W & \approx & \frac{\alpha^\prime}{\alpha^\prime + \alpha_2}[1 +\frac{4 \pi \alpha_2}{\alpha^\prime + \alpha_2}(-\alpha^\prime b^\prime + \alpha_2 b_2) \ln ({M_{Z}^2}/{M_M^2})] \nonumber \\
&\approx &(1/4)[1 + 4\pi \alpha_2(-\frac{1}{4}b^{\prime}+\frac{3}{4}b_2) \ln(M_Z^2/M_M^2)]  \, ,
\eea
where
\be
\label{b2}
b_2=(1/16\pi^2)[\frac{22}{3}-\frac{4}{3}F-\frac{1}{6}n_H-\frac{2}{3}n_3]
\ee 
and
\be
\label{bprime}
b^\prime =-(1/16\pi^2)[\frac{20}{9}F+\frac{1}{6}n_{H}+\bar{n}_3] \, .
\ee
The scalar contributions to Eqs.~(\ref{b2},\ref{bprime}) come from $-(1/3)T_S$ with $T_S=1/2,2$ (doublets and triplets) for $b_2$ and $(1/3)\sum (Y_S/2)^2$ with $Y_S/2=1/2, 1$ (doublet and complex triplet) for $b^\prime$. 
Tabulated below are some examples of spectra with $M_M$ in the range (\ref{monmass2}) that yield $0.230 < x_W < 0.233$,
to be compared with the experimental central value $x_W = 0.23121$ in the $\overline{MS}$ prescription~\cite{PDG} (one should
allow for higher-order uncertainties in the renormalization calculation).\\
\begin{center}
\begin{tabular}{|c| c| c| c| c| c|}
\hline
 $M_M$ (TeV) & $F$ & $n_H$ & $n_3$ & $\bar{n}_3$ & $x_W$  \\ 
\hline
2.3 & 3 & 1 & 0 & 0 & 0.232 \\
\hline
3 & 3 & 3 & 0 & 0 & 0.2314 \\
\hline
3 & 3 & 1 & 1 &1 & 0.2318 \\
\hline
3 & 4 & 1 & 0 & 0 & 0.2328 \\
\hline

\end{tabular}
\end{center}

We 
note that cases with $F=5, 6$ are disfavoured experimentally, 
as they yield ${\sin}^2\theta_W(M_Z^2) > 0.233$. Also disfavoured are scenarios such as $n_H =2$, $n_3 = 1$, $\bar{n}_3 = 1$ light Higgs fields below $M_M$.

The EW-$\nu_R$ model we have studied here has many interesting properties. In addition to containing a seesaw scenario for 
neutrino masses that predicts several possibilities for new particles that could be detected at the LHC, it also predicts the
existence of an electroweak magnetic monopole with mass $\lesssim 3$ TeV, light enough to be detected in principle by the MoEDAL experiment~\cite{moedal, mm2}.

Remarkably,
as we have shown in this paper, the Dirac Quantization Condition needed for the quantum consistency of the EW-$\nu_R$
model imposes a specific value of the weak mixing parameter $\sin^2 \theta_W = 1/4$ at the monopole mass scale.
Plausible choices of the monopole mass and the numbers of fermions and Higgs bosons with masses below that of the
monopole yield predictions for the renormalized weak mixing parameter ${\rm sin}^2\theta_W(M_Z^2)$ that are consistent
with experimental measurements, within the theoretical uncertainties.
The success of this prediction has interesting implications on the Majorana masses of right-handed neutrinos,
since both quantities depend on the Higgs triplet VEV $v_M$, as well as the spectra of light new particles. With Majorana masses of the EW-$\nu_R$ model being $M_R = g_{\nu_R} v_M$, Eq.~(\ref{monmass}) gives an interesting relation between the monopole and right-handed neutrino Majorana masses $M_M= \frac{4\pi}{g\, g_{\nu_R}} f(\lambda/g^2) M_R$. 
Electroweak-scale non-sterile $\nu_R$ could be discovered via like-sign dilepton events with displaced vertices, and give a range for the monopole mass: $19 \, M_R \lesssim M_M  \lesssim 34 \, M_R$ for $f(\lambda/g^2) =1,1.78$, with $g \sim 0.65$ and assuming $g_{\nu_R} \sim 1$. The search for charged mirror quarks and leptons, which are long-lived particles in this model, have been discussed in \cite{pqnur}.
A detailed analysis of this and other aspects of the model are beyond the scope of this article and will be given elsewhere.

\vspace{0.1cm}

\begin{acknowledgments}

We thank T. Vachaspati for discussions. 
The work of J.E. and NEM and is supported in part by the UK Science and Technology Facilities (STFC) under the research grant ST/P000258/1, and J.E. is also  supported in part by an Estonian Research Council Mobilitas Pluss grant. J.E. and N.E.M. participate in the COST Association Action CA18108 ``{\it Quantum Gravity Phenomenology in the 
Multimessenger Approach (QG-MM)}''. NEM also acknowledges a scientific associateship (``\emph{Doctor Vinculado}'') at IFIC-CSIC, 
Valencia University, Valencia, Spain.

\end{acknowledgments}


\begin{thebibliography}{99}
\bibitem{Mirror}
 The heavy mirror fermions with chiralities opposite to SM fermions are present because of the global symmetries U(1)$_{\rm SM} \times$ U(1)$_{\rm MF}$ needed for the seesaw mechanism in~\cite{pqnur}.
\bibitem{Alternative}
An alternative way to understand the necessity of \eqref{pred2}
stems from the fact that, if one uses \eqref{egrel}, then from the long-distance expression (\ref{Bnew}) for the 
$\gamma-Z$ magnetic field, one observes that the magnetic charge of this field is 
$\tilde g = \frac{{\rm sin}^2\theta_W}{e} = \frac{{\rm sin}\theta_W}{g}$, i.e., $g \, \tilde g = {\rm sin}\theta_W$. Comparing with 
the quantization condition \eqref{quantize}, we then obtain \eqref{pred2}.
\bibitem{mm2}  Production mechanisms and potential suppression due to monopole structure~\cite{drukier} are reviewed in 
\cite{mm}.
\end{thebibliography}
\end{document}